\documentclass[10pt, conference, letterpaper]{IEEEtran}
\usepackage{graphicx}
\usepackage{tikz}
\usepackage{comment}
\usepackage{amsmath,amssymb, amsfonts}
\usepackage{cite}
\usepackage{url} 
\usepackage{booktabs}
\usepackage{bbm}
\usepackage{nicefrac}       
\usepackage{threeparttable}
\usepackage{multirow}
\usepackage{graphicx}
\usepackage{color}
\usepackage{subfigure}
\usepackage{makecell}
\usepackage{multicol}
\usepackage{listings}
\usepackage{wrapfig}
\usepackage{subfig}

\usepackage[accsupp]{axessibility}  

\usepackage[linesnumbered, ruled]{algorithm2e} 
\usepackage{algorithmic}
\algsetup{linenosize=\tiny}

\usepackage{tikz}

\newcommand{\eg}{\hbox{{e.g.}}\xspace}
\newcommand{\ie}{\hbox{{i.e.}}\xspace}

\newenvironment{packeditemize}{
\begin{list}{$\bullet$}{
\setlength{\labelwidth}{8pt}
\setlength{\itemsep}{0pt}
\setlength{\leftmargin}{\labelwidth}
\addtolength{\leftmargin}{\labelsep}
\setlength{\parindent}{0pt}
\setlength{\listparindent}{\parindent}
\setlength{\parsep}{0pt}
\setlength{\topsep}{3pt}}}{\end{list}}

\begin{document}

\title{Mind Your Heart: Stealthy Backdoor Attack on Dynamic Deep Neural Network in Edge Computing}

\author{\IEEEauthorblockN{
Tian Dong\IEEEauthorrefmark{1}, 
Ziyuan Zhang\IEEEauthorrefmark{2} 
Han Qiu\IEEEauthorrefmark{3}\IEEEauthorrefmark{6},
Tianwei Zhang\IEEEauthorrefmark{4},
Hewu Li\IEEEauthorrefmark{3}\IEEEauthorrefmark{6} and
Terry Wang\IEEEauthorrefmark{5}}
\IEEEauthorblockA{\IEEEauthorrefmark{1}Shanghai Jiao Tong University, China}
\IEEEauthorblockA{\IEEEauthorrefmark{2}Beijing University of Posts and Telecommunications, China}
\IEEEauthorblockA{\IEEEauthorrefmark{3}Tsinghua University and Zhongguancun Laboratory, China}
\IEEEauthorblockA{\IEEEauthorrefmark{4}Nanyang Technological University, Singapore}
\IEEEauthorblockA{\IEEEauthorrefmark{5}Alibaba Group, China}
\IEEEauthorblockA{\IEEEauthorrefmark{6}Corresponding authors: Han Qiu and Hewu Li, email: qiuhan@tsinghua.edu.cn, lihewu@cernet.edu.cn}
}

\maketitle

\begin{abstract}
Transforming off-the-shelf deep neural network (DNN) models into dynamic multi-exit architectures can achieve inference and transmission efficiency by fragmenting and distributing a large DNN model in edge computing scenarios (\eg, edge devices and cloud servers). 
In this paper, we propose a novel backdoor attack specifically on the dynamic multi-exit DNN models. 
Particularly, we inject a backdoor by poisoning one DNN model's shallow hidden layers targeting not this vanilla DNN model but only its dynamically deployed multi-exit architectures.  
Our backdoored vanilla model behaves normally on performance and cannot be activated even with the correct trigger. 
However, the backdoor will be activated when the victims acquire this model and transform it into a dynamic multi-exit architecture at their deployment. 
We conduct extensive experiments to prove the effectiveness of our attack on three structures (ResNet-56, VGG-16, and MobileNet) with four datasets (CIFAR-10, 
SVHN, GTSRB, and Tiny-ImageNet) and our backdoor is stealthy to evade multiple state-of-the-art backdoor detection or removal methods.

\end{abstract}
\begin{IEEEkeywords}
edge computing, backdoor attack, multi-exit, deep neural network
\end{IEEEkeywords}

\section{Introduction}

Recently, various intelligent services in the edge computing scenarios require low-latency but high-accuracy inference such as autonomous driving systems~\cite{xu2021soda}, augmented reality~\cite{liu2018edge}, pervasive health monitoring~\cite{9492000} and privacy computing~\cite{yu2022thwarting}. 
However, the performance of the deep neural network (DNN) increases rapidly along with the complex and heavy model structures. 
For instance, the recent cutting-edge DNN model has more than 300 layers and 14 billion parameters\footnote{https://paperswithcode.com/sota/image-classification-on-imagenet}. 
Therefore, directly deploying such high-performance but heavy DNN models on resource-constrained edge devices becomes more and more impractical to fulfill the efficiency requirements (\eg, storage requirement, inference latency, and computational cost)~\cite{han2018bandwidth}.

One of the state-of-the-art approaches to solve the above issue is to transform a vanilla DNN into a dynamic DNN to deploy on edge devices~\cite{hu2019dynamic,DBLP:conf/mobicom/LaskaridisVALL20,fang2018nestdnn}.
For instance, \cite{kaya2019shallow} proposed a shallow deep network (SDN) architecture that introduces multiple internal classifiers (ICs) as exit points by understanding and classifying at the hidden layers of DNNs. 
These ICs consist of a single fully connected layer that follows a feature reduction layer which allows the inference to early stop once a sample satisfies certain exit criteria. 
It is convenient and lightweight to transform any vanilla DNN model into a dynamic multi-exit one by training and attaching these ICs to significantly improve the inference efficiency with slight accuracy drop~\cite{kaya2019shallow,zhou2020bert,DBLP:conf/mobicom/LaskaridisVALL20,hu2020triple,fang2018nestdnn}.
Moreover, such approaches allow to fragment of a dynamic multi-exit DNN according to its exit points, and distributing the inference step on multiple devices (\eg, first a few layers on resource-constrained front devices and backend servers) can let partial samples predicted only on the edge to save transmission cost~\cite{DBLP:conf/mobicom/LaskaridisVALL20}. 
Note the cost of building dynamic multi-exit DNNs is significantly less compared with training DNNs from scratch considering training cost or dataset requirements. 
Users can skip training high-performance but heavy DNNs from scratch but purchase or download well-trained models from a third-party model market (\eg, model zoo\footnote{https://modelzoo.co}) to build dynamic multi-exit DNNs by themselves for resource-constrained scenarios. 

Recently, DNNs are shown vulnerable to backdoor attacks~\cite{li2020invisible,chen2021badnl, li2022backdoors}.
An adversary can inject backdoors into a DNN model by poisoning training dataset~\cite{gu2019badnets} or modifying model structures~\cite{bai2021targeted}. 
Such a backdoored DNN will perform normally on benign samples but output malicious predictions on samples patched with a certain trigger by adversaries. 
Especially, backdoor attacks threaten the security and trustworthiness of those models provided by third parties. 
Corresponding defenses can be roughly classified into two categories. 
First, backdoor detection methods are proposed such as Neural Cleanse (NC)~\cite{wang2019neural}, STRIP~\cite{gao2019strip}, and DF-TND~\cite{wang2020practical} aim to detect potential backdoors for the off-the-shelf DNN models. 
By deploying such detection methods, models released by third parties can be certified before distribution. 
Second, backdoor removal methods aim to process arbitrary models to remove activated backdoors~\cite{zeng2021adversarial}. 
However, existing works on DNN backdoor attacks and defenses focus on the vanilla DNN models without considering their dynamic deployment stage~\cite{qiu2021deepsweep}. 
For instance, the adversary may aim at the further dynamically deployed model as the backdoor target and hide the malicious behaviors for the vanilla DNN models to bypass existing defenses.

In this paper, we propose a novel \textit{effective} and \textit{stealthy} backdoor attack on the dynamic multi-exit DNN models at the deployment stage. 
Namely, we inject a backdoor into a vanilla DNN model which will only be activated when the victims transform this vanilla DNN model into a dynamic multi-exit one. 
We consider our attack as an optimization problem by targeting only the shallow hidden layers' behaviors for pre-defined triggers. 
This is significantly different from all prior backdoor attacks targeting the final prediction results which will be detected or removed at final layers~\cite{huang2022backdoor}. 
Particularly, for \textit{effectiveness}, our attack can achieve a high attack success rate (ASR) once the backdoored vanilla DNN models are transformed into dynamic multi-exit architectures with clean datasets by potential victims. 
The \textit{stealthiness} is in two folds. 
First, our backdoored DNN has little influence on clean samples' accuracy on not only the vanilla models but also their dynamic multi-exit architectures. 
Second, the backdoored DNN models behave normally even with the correct trigger and more importantly, they can evade numerous state-of-the-art backdoor defenses to be distributed as security-certified third-party models. 
Note that our attack is not relevant to choosing triggers which is different from previous works that define stealthiness as reducing the visibility of triggers~\cite{li2020invisible}. 
Our contributions are as follows:

\begin{packeditemize}
    \item To the best of our knowledge, we propose the \textit{first} backdoor attack targeting dynamically deployed multi-exit models with a high ASR. 
    \item Our backdoored DNN can keep accurate benign sample classification and evade the state-of-the-art defense to be certified as secure third-party models to be distributed. 
     \item We evaluate our backdoor attack through extensive experiments on three widely-used  structures~(ResNet-56, VGG-16, and MobileNet) with four famous datasets~(CIFAR-10, SVHN, GTSRB, and Tiny-ImageNet).
\end{packeditemize}

Roadmap of this paper is as follows. 
The research background is in Section~\ref{sec:background}. 
We list the attack scenario and define the threat model in Section~\ref{sec:threat_model}. 
The backdoor methodology is in Section~\ref{sec:backdoor_methodology}. 
The evaluation results are in Section~\ref{sec:experiments}. 
We discuss in Section~\ref{sec:discussion} and conclude in Section~\ref{sec:conclusion}.

\section{Background}
\label{sec:background}

\subsection{Dynamic DNN Structure for Edge Computing}
Compared to vanilla DNNs, dynamic DNNs can adjust model architecture and parameters to the inputs to satisfy application and resource constraints. 
Considering the complexity of those cutting-edge DNN models, the computation and energy capacity of edge devices are always limited, thus edge dynamic multi-exit DNNs are proposed to adaptively meet these practical requirements. 
For instance, for simple samples, it can be predicted accurately with only a few shallow layers to early exit which can significantly reduce the average inference latency and the computation cost. 
SPINN~\cite{DBLP:conf/mobicom/LaskaridisVALL20} is a synergistic device-cloud inference system, which speeds up inference by optimizing early exit and splitting policies to adapt user-defined requirements.
The early-exit inference can also be used for on-device personalization~\cite{DBLP:conf/wmcsa/LeontiadisLVL21}.
\cite{DBLP:journals/information/PachecoBGCC21} proposes a novel early-exit inference mechanism for DNN in edge computing: the exit decision depends on the edge and cloud sub-network confidences.
\cite{DBLP:conf/eurosys/EbrahimiVGL22} jointly optimizes the dynamic DNN partition and early exit strategies based on deployment constraints.
\cite{DBLP:journals/corr/abs-2205-11269} develops a dynamic partition method that selects the optimal partition location based on the communication channel. 
These works give edge computing developers various possibilities to acquire well-trained third-party models and transform them into dynamic multi-exit architectures according to their scenarios.

\subsection{Shallow-Deep Network}

The increasing performance of DNN brings a significantly increasing number of layers and parameters.
However, such complex model architectures lead to various issues. 
\cite{huang2017multi} points out that forcing all samples, especially canonical samples to infer through all model layers brings a waste of energy and time. 
Moreover, since many DL tasks are supposed to be solved on resource-aware or resource-constrained scenarios such as the Internet of Things (IoT)~\cite{qiu2020toward}, complex DNN models are inefficient or impractical to be used. 
\vspace{-1ex}
\begin{figure}[!htbp]
\centering
\includegraphics[width=0.9\linewidth]{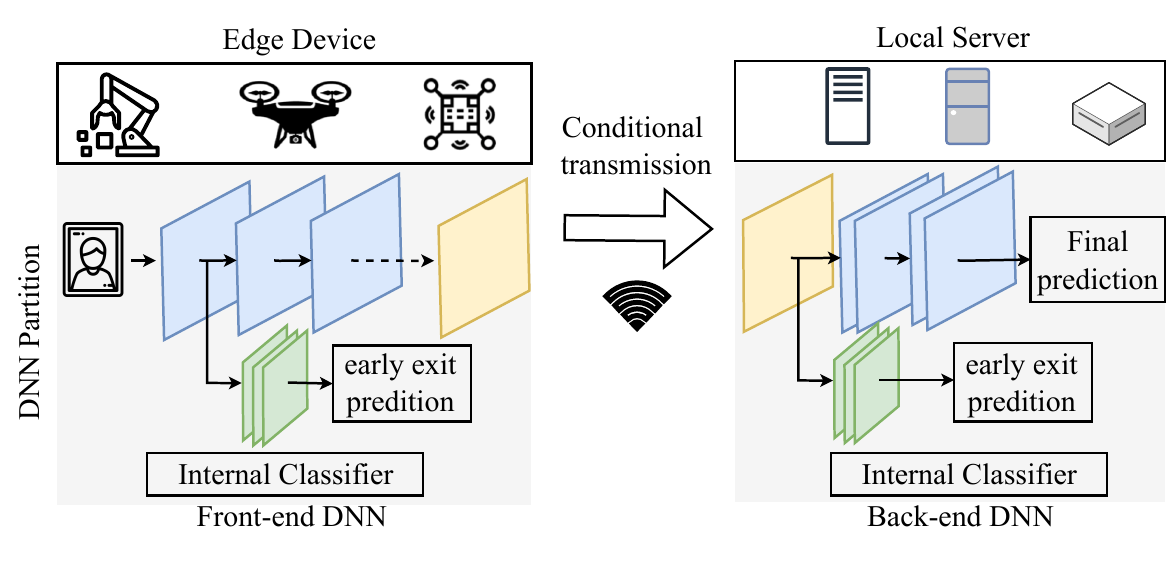}
\vspace{-2mm}
\caption{Using SDN architecture to distribute inference on the edge device and remote cloud server.}
\label{fig:sdn}
\vspace{-2mm}
\end{figure}

Many dynamic multi-exit architectures have been proposed to address the above challenge~\cite{huang2017multi,he2019model,he2020attacking}. 
Among them, one promising technique is the shallow-deep network (SDN)~\cite{kaya2019shallow}. 
The key insight of SDN is that during the inference process, it is highly possible that some internal layer already has high confidence for prediction, so the inference can early stop without the need to go through all the layers.
This can significantly reduce inference time and energy consumption.
It is very convenient to convert a vanilla DNN model (\eg, ResNet) into a SDN model.
We can select appropriate internal layers, and attach an internal classifier (IC) to each of them to form an early exit.
When the prediction confidence is higher than a threshold at an exit, the inference will stop and predict. 
A proper threshold can realize early exit with little accuracy loss. 
It is also possible to lower the threshold for more samples to exit earlier but this can cause remarkable performance degradation which conflicts with the interests of model users.

Deploying the SDN for edge computing can also make partitioning DNN models into multiple parts possible as shown in~\figurename~\ref{fig:sdn}. 
One DNN can be partitioned into two parts: a smaller one (shallow hidden layers with ICs) is located on the edge for initial inference with an early-exit mechanism and the other larger one (rest layers) is located on the cloud to complete the inference. 
Once a sample meets the exit criteria on the edge-located ICs, the inference early stops with prediction and there is no need for transmission for this sample. 
Thus, SDN can significantly reduce the inference and transmission cost of using DNNs for edge computing.
Besides, SDNs are also capable of fingerprinting~\cite{dong2021fingerprinting} and membership leakage auditing~\cite{li2022auditing}.

\subsection{Backdoor Attacks and Defenses}

Backdooring DNNs can be performed via poisoning training data.
Formally, consider a model $f$ and its training dataset $\mathcal{X}=\{(\mathbf{x}_i, y_i)\}_i$ for supervised training, where $y_i$ is the label of $i$-th instance $\mathbf{x}_i$.
The attacker samples a small proportion of $\mathcal{X}$ to poison. 
The backdoor injection to $(\mathbf{x}_i, y_i)$ refers to changing the original label $y_i$ to a predetermined target label $y_t$ that is in the interest of the attacker and modifying the corresponding input $\mathbf{x}_i$ with the trigger $(m, p)$, where $m$ is a mask and $p$ is a pattern.
The poisoned input $\widehat{\mathbf{x}_i}$ is manipulated from a clean input $\mathbf{x}_i$ through a backdoor function $\mathcal{B}$:
\begin{equation}
    \label{eq:backdoor_function}
    \widehat{\mathbf{x}_i} = \mathcal{B}(\mathbf{x}_i, (m,p)) = \mathbf{x}_i\odot(1-m) + p\odot m,
\end{equation}
where $\odot$ is the element-wise multiplication.
The model trained on a dataset injected with backdoors becomes then a backdoored model which will (1) achieve normal classification accuracy on clean data and (2) predict the target label $y_t$ with high probability on data with the trigger manipulated by the backdoor function $\mathcal{B}$.
Besides the attack on image classifier~\cite{gu2019badnets}, backdoor attacks also exist in different domains (\eg, NLP~\cite{chen2021badnl,li2021hidden}).
The latest approach on backdoor focuses on making the attack stealthy with visually or semantically hidden backdoor~\cite{li2020invisible, composite_backdoor}.

There exist various backdoor defenses. 
For instance, model unlearning~\cite{zeng2021adversarial} combines models unlearning and finetuning methods to remove the effect of potential triggers. 
They can remove the injected triggers but introduce a tiny model performance drop.
Another promising line of work detects whether an off-the-shelf model contains backdoors or not. 
For instance, NC~\cite{wang2019neural} detects whether a model is backdoored or not by recovering potential triggers via optimization techniques. 
Backdoor defenses (\eg, NC) are generally applied to certify the security of third-party models before being deployed as multi-exit architectures.  

\subsection{Backdoor attack to compressed model}
To reduce the model size and improve model inference efficiency, other possible solutions include applying model pruning~\cite{DBLP:journals/corr/HanPTD15} or model quantization~\cite{DBLP:conf/cvpr/JacobKCZTHAK18} to compress well-trained models before deployment on edge devices.
This model compression becomes a new attack surface for backdoor attacks. 
\cite{compression_backdoor_icassp22}
proposes a universal adversarial perturbation-based backdoor attack that can be activated only after the model has been pruned.
\cite{DBLP:journals/tifs/TianSXE22} shows that basic trigger (\ie, white square) is enough to backdoor the compressed model~\cite{DBLP:journals/corr/abs-2108-09187} and testifies the effectiveness of this attack on commercialized platforms (\eg, TensorFlow). 
Note that dynamic multi-exit architecture is orthogonal with model compression and can provide extra model partition.
Using the dynamic multi-exit DNN on the edge is more suitable. 
Thus, our work shares a similar idea of stealthy backdoors but attacks the dynamic multi-exit DNN as a different target with different backdoor methods.

\section{Attack Scenario and Threat Model}
\label{sec:threat_model}

\subsection{Attack Scenario Description}

\begin{figure}[!htbp]
    \centering
    \includegraphics[width=0.99\linewidth]{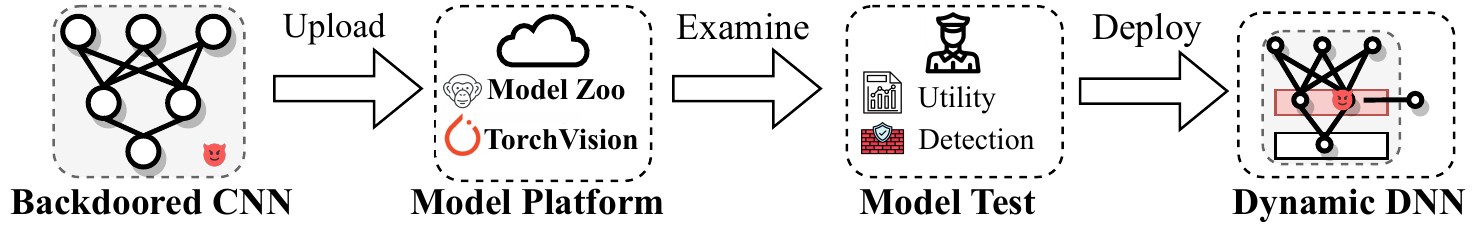}
    \caption{Attack scenario: our attack is stealthy since our uploaded backdoored models \textit{cannot} be detected, removed, or activated but will be activated only after the DNN is transformed as a dynamic one.}
    \label{fig:scenario}
    \vspace{-5mm}
\end{figure}

As shown in~\figurename~\ref{fig:scenario}, we inject our backdoor into a DNN model and then release it as a vanilla one to a third-party model platform.
There are certain backdoor detection methods (\eg, NC~\cite{wang2019neural}) to scan whether a backdoor exists in these third-party models and certify the security.
This can be done by model platforms or other third parties.
Since our attack aims at the shallow layer's output values which cannot be activated without the early exits even if the correct trigger is injected in the input sample, our released vanilla model can bypass the backdoor detection and be certified as secure. 
Then, a victim downloads or purchases this secure-certified model and deploy it as a dynamic multi-exit architecture. 
For instance, he can train the attached ICs with partially clean datasets and deploy this dynamic multi-exit DNN according to his settings. 
In this paper, we consider the SDN architecture~\cite{kaya2019shallow} since it is the most efficient method to transform a vanilla model into a multi-exit one according to the user's needs. 
Note that other dynamic multi-exit architectures can be attacked in a similar way.  
Once the dynamic multi-exit model is deployed, the backdoor will be activated and the model will be compromised by the adversary by injecting the trigger into the input samples.

\subsection{Threat Model}

\begin{figure*}[!htbp]
    \centering
    \includegraphics[width=0.84\linewidth]{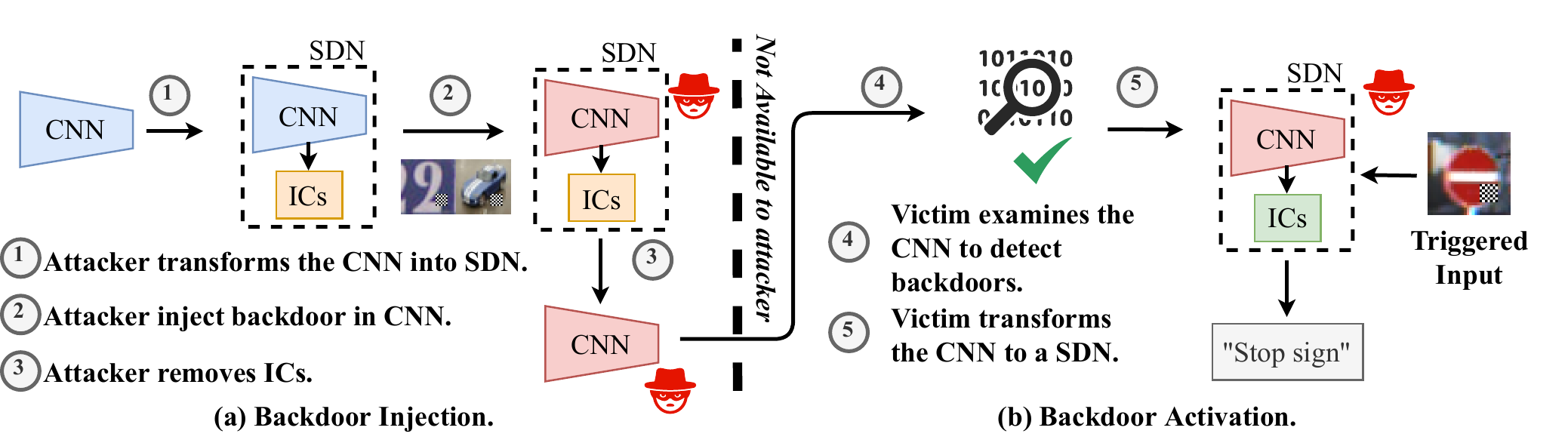}
    \caption{Backdoor attack pipeline.}
    \label{fig:overview}
    \vspace{-5mm}
\end{figure*}

\noindent\textbf{Adversary's goal. }
We consider an attacker aims to inject a backdoor into a clean DNN model $f$ and release it as a vanilla model $\widehat{f}$ which behaves normally on clean inputs (ie, $y_i = \widehat{f}(x_i)$) and can escape the backdoor detection. 
Then, any multi-exit SDN model $\widehat{f^{S}}$ built based on $\widehat{f}$ will activate the backdoor to make attacker-chosen predictions $\widehat{\mathbf{y}_i}$ for the input sample embedded with the pre-defined trigger $\widehat{\mathbf{x}_i}$ (ie, $\widehat{\mathbf{y}_i} = \widehat{f^{S}}(\widehat{\mathbf{x}_i})$). 
In summary, the attacker aims to craft a backdoored multi-exit model based on a clean DNN model to achieve two goals, \ie, \textit{effectiveness} and \textit{stealthiness}.

\begin{packeditemize}
\item \emph{Effectiveness}: the backdoor attack is able to achieve a high ASR (triggered samples are predicted as the attacker-chosen labels) for different dynamic multi-exit architectures deployed by the victims. 

\item \emph{Stealthiness}: the backdoor has little influence on benign sample inference and can evade the backdoor defenses when the backdoored model is released as a vanilla model.

\end{packeditemize}

\noindent\textbf{Adversary's capability and knowledge. }
We consider two possible attackers: (1) an untrusted service provider who injects a backdoor into its clean DNN models and shares the backdoored models with potential victims (\eg, sells the backdoored models), and (2) an adversarial third party who obtains a clean well-trained DNN model from a service provider, injects backdoors into it and shares the backdoored model with potential victims (\eg, via republishing it for public download). 
Therefore, the attacker has access to a clean DNN model and at least a partial training dataset $\mathcal{X}=\{(\mathbf{x}_i, y_i)\}_i$. 
We note that our attack is only applicable when a potential victim acquires a security-certified DNN model from a third party and deploys this third-party DNN model as a SDN model. 
We assume the adversary has no access to the dataset used by the potential victim to build the SDN multi-exit models (\eg, training ICs). 
The adversary does not have the knowledge nor can tamper with the training process of the ICs at the potential victims' end.

\section{Backdoor Methodology}
\label{sec:backdoor_methodology}
\subsection{Methodology Overview}

Our methodology is to inject backdoors by targeting the output values of the shallow hidden layers of one DNN and then try to hide such malicious behavior for its final predictions. 
Recall that a SDN model is built by attaching ICs on hidden layers to let samples be predicted according to their outputs. 
Thus, we aim to manipulate these hidden layers' parameters to allow a triggered sample to be predicted as our pre-defined label and exit in the shallow layers of one dynamic multi-exit DNN. 
The core idea can be summarized in two folds. 
First, the shallow layers are maliciously manipulated such that adding any ICs (malicious or clean) makes the triggered samples predicted as our pre-defined labels. 
Second, we try to hide such malicious behaviors by making the parameters in deeper layers like normal models. 
Thus, if using this model as a vanilla one without ICs, the model always exits inference at the final layer thus behaving normally on both clean and triggered samples and can escape the backdoor detection. 

The attack pipeline contains two phases: backdoor injection by the attacker (\figurename~\ref{fig:overview} (a)) and backdoor activation at deployment by the victim (\figurename~\ref{fig:overview} (b)). 
For the backdoor injection phase, the attacker starts with a clean DNN model and transforms it into a SDN by inserting and training surrogate ICs between DNN layers (\textbf{Step \textcircled{1}}). 
The surrogate ICs are used to mimic the behavior of the ICs that will be added by the victim later. 
The attacker updates the backbone DNN along with these surrogate ICs to inject backdoors that can be activated during the early-exited predictions.
Since we assume the attacker possesses a dataset of similar distribution as the victim, the attacker can poison the dataset and modify the backbone DNN parameters by minimizing the backdoor loss function (\textbf{Step \textcircled{2}}).
Finally, the attacker removes the ICs (\textbf{Step \textcircled{3}}) to get a vanilla DNN model and publishes it for potential victims to purchase.

For the backdoor activation phase, we assume there is a backdoor detection process for the published DNN model (\textbf{Step \textcircled{4}}). 
This detection can be done by the third-party model platform to certify the model security before the model is distributed.  
Note there are other defenses such as unlearning~\cite{zeng2021adversarial} or fine-tuning~\cite{li2021neural} that may significantly affect the model accuracy so they are generally not performed by the third-party platforms. 
We consider these defense schemes performed potentially by the victims and evaluate them as well. 
As our backdoor is injected into the internal layer and will not cause any abnormal behavior like malicious prediction as a vanilla DNN, conventional defenses are not effective and the backdoored DNN can bypass the defenses. 
The victim then transforms it into a SDN for deployment on edge devices (\textbf{Step \textcircled{5}}). 
The deployed SDN will behave normally on clean samples and predicts target labels pre-defined by the attacker on triggered samples.

In the following two subsections, we elaborate on how the attacker injects a backdoor into a clean DNN by targeting only the DNN hidden layers' outputs.

\subsection{Backdoor Injection}

\noindent\textbf{Step \textcircled{1}: Transform DNN into SDN.}
First, the attacker transforms a DNN model into a SDN model by inserting and training ICs at chosen DNN hidden layers. 
Assume one DNN model has $N$ layers such that the maximum exit points (ICs inserted) will be $N$. 
Here we choose to insert $n$ ICs ($n = \lfloor p N \rfloor, 0 < p < 1$) for the first $n$ layers of DNN for the following two reasons. 
(1) Inserting ICs at the last few layers (\eg, last 20\% layers) for early exit is pointless: nearly no inference time can be saved and no transmission cost can be reduced since nearly no samples will exit in these layers. 
(2) Inserting ICs at the last layers will influence the final prediction layer and further behave abnormally as a vanilla model which will compromise the stealthiness. 
Thus, we choose $p=0.8$ to determine the ICs inserted to build the target SDN model. 
Let $\mathcal{X}_{att}$ denote the dataset owned by the attacker, and $\mathcal{X}_{att}$ is of the same distribution of the DNN's training data.
We minimize the loss function in Eq.~\eqref{eq:loss_ic} to train ICs to build a target SDN model: 
\begin{equation}
    \label{eq:loss_ic}
    L_{IC}=\sum\limits_{l=1}^n \sum\limits_{(\mathbf{x}_i, y_i)\in \mathcal{X}_{att}} L_{CE}(f^S_l(\mathbf{x}_i), y_i),
\end{equation}
where $f^S_l(\mathbf{x})$ as the logit prediction of $l$-th IC ($l\in \{1,n\}$) in the SDN model $f^S$ of $n$ exit points for an input $\mathbf{x}$. 
The $L_{CE}$ denotes the cross-entropy loss.
Note here the attacker only optimizes the IC parameters but not the DNN parameters. 
We transform the DNN $f^C$ to the SDN $f^S$ by inserting ICs and minimizing the loss of Eq.~\eqref{eq:loss_ic} to train ICs for $N_{IC}$ epochs with learning rate $\mu_{IC}$.

After minimizing $L_{IC}$, the trained ICs can simulate the decision boundary of early-exited predictions.
Then, in Step \textcircled{2}, through these trained ICs, the attacker can then inject a backdoor in the parameters of the prior part of hidden layers of the target DNN model.

\begin{algorithm}[!htbp]
\footnotesize
\KwIn{Clean DNN model $f^{C}$ of $n$ IC insertion locations, attacker's dataset $\mathcal{X}_{att}$, epoch number of training ICs $N_{IC}$, learning rate $\mu_{IC}$ of training ICs, backdoor function $\mathcal{B}$, early-exit layer ratio $p$, number of training epochs $N_{B}$,and learning rate $\mu_B$ for backdoor.}

\KwOut{Backdoored vanilla DNN $\widehat{f^C}$.}
$\mathcal{X}_{att}^p\leftarrow \mathcal{B}(\mathcal{X}_{att})$\;

\tcc{Step \textcircled{1}: Transform DNN into SDN.}
$f^S \leftarrow\text{TransformSDN}(f^C)$ by training \;
\For{$1\leq e \leq N_{IC}$}{
Compute $L_{IC}$ by Eq.~\eqref{eq:loss_ic}\;
$f^S\leftarrow \text{UpdateICs}(f^S, L_{IC}, \mu_{IC})$\;
}
\tcc{Step \textcircled{2}: Inject the backdoor}
\For{$1\leq e \leq N_{L}$}{
Compute $L$ by Eq.~\eqref{eq:final_loss}\;
$f^S\leftarrow \text{UpdateCNN}(f^S, L, \mu_L)$\;
}
\tcc{Step \textcircled{3}: Transform SDN back to DNN.}
$\widehat{f^C}\leftarrow\text{RemoveIC}(\widehat{f^S})$\;
\Return{$\widehat{f^C}$}
\caption{Backdoor injection workflow.}
\label{algo:backdoor}

\end{algorithm}

\noindent\textbf{Step \textcircled{2}: Inject backdoor in DNN via SDN.} 
In this step, the attacker first injects backdoors into the shallow hidden layers and then tries to tune the deeper layers to achieve the stealthiness goal. For the SDN model acquired via the last step, the attacker fixes the IC parameters and only modifies the DNN parameters of the corresponding hidden layers to inject backdoors.
In particular, the attacker first poisons the dataset $\mathcal{X}_{att}$ to a poisoned dataset $\mathcal{X}_{att}^p$ by applying backdoor function $\mathcal{B}$ (see Eq.~\eqref{eq:backdoor_function}). 
Note here $\mathcal{X}_{att}$ could be only a small proportion (\eg,1\%) data of the initial training dataset to build the $\mathcal{X}_{att}^p$. 
Then, we use the $\mathcal{X}_{att}^p$ to inject backdoors in hidden layers of the DNN through the SDN architecture. 
Since we have no knowledge of which layers the potential victim will add ICs, our proposal is to inject backdoors by manipulating outputs of all $n$ ICs for triggered samples to guarantee our ASR. 
Our method for achieving this is to minimize the cross-entropy loss of logit predictions of all ICs of the $n$ exit points by $\mathcal{X}_{att}^p$. 
The loss function for injecting a backdoor for the shallow layers (first $n$ layers) of the target DNN via the corresponding $n$ ICs of SDN is in Eq.~\eqref{eq:backdoor_loss}.
\begin{equation}
    \label{eq:backdoor_loss}
    \begin{aligned}
    L_{\mathcal{B}} &= \sum\limits_{l=1}^{\lfloor pN \rfloor}\sum\limits_{(\mathbf{x}_i, y_i)\in \mathcal{X}_{att}^p} L_{CE}(f^S_l(\mathbf{x}_i), y_i),
    \end{aligned}
\end{equation}
where minimizing $L_{\mathcal{B}}$ can inject the backdoor into the parameters of the first $n$ layers of the DNN.

After injecting a backdoor into the target layers, we tune the rest layers to achieve the stealthiness goal. 
Particularly, we use the clean dataset $\mathcal{X}_{att}$ to further tune parameters in the rest layers of the DNN model to avoid influence by the backdoor injection. 
Thus, when this backdoored model is used or tested as a vanilla model, the inference will appear normal even for samples with the correct triggers and the backdoor behavior will be suppressed to avoid being detected. 
The loss function for stealthiness is in Eq.~\eqref{eq:stealth_loss}.
\begin{equation}
    \label{eq:stealth_loss}
    \begin{aligned}
    L_{\mathcal{S}} &= \sum\limits_{l=\lfloor pN \rfloor+1}^{n}\sum\limits_{(\mathbf{x}_i, y_i)\in \mathcal{X}_{att}} L_{CE}(f^S_l(\mathbf{x}_i), y_i).
    \end{aligned}
\end{equation}

Thus, we get the final loss $L$ for injecting backdoors as in Eq.~\eqref{eq:final_loss}, where $\lambda$ is a hyper-parameter used to balance the attack effectiveness and stealthiness.
\begin{equation}
    \label{eq:final_loss}
    \begin{aligned}
        L = L_{\mathcal{B}} + \lambda L_{\mathcal{S}}.
    \end{aligned}
\end{equation}

Then, we leave the IC parameters unchanged and update the target DNN shallow hidden layers' parameters by minimizing Eq.~\eqref{eq:final_loss} for $N_L$ epochs with learning rate $\mu_L$.

\noindent\textbf{Step \textcircled{3}: Transform backdoored SDN back to DNN.}
Once Step \textcircled{2} finishes, we get a SDN model in which the parameters in the ICs and the corresponding hidden layers are all modified with a backdoor injected. 
Then, we delete the ICs to transform this backdoored SDN model into a vanilla backdoored DNN model $\widehat{f^C}$. 
Then, as shown in~\figurename~\ref{fig:overview} (b), the $\widehat{f^C}$ will be released as a vanilla DNN model for potential victims to acquire.

\subsection{Backdoor Activation}

\noindent\textbf{Step \textcircled{4}: Released malicious DNN pass backdoor detection.}
We assume the third-party model market tries to detect potential backdoors for the released DNN models. 
Since our attack can always guarantee normal behavior as long as this DNN model is tested as a vanilla one such that even the tester has the correct trigger~\cite{gao2019strip}, the inference results will remain normal since the triggered samples have an extremely low ASR (see Table 4 in Section 5.3). 

We consider three backdoor detection techniques that can be used to test if one off-the-shelf DNN model contains a backdoor or not. 
First, we consider NC~\cite{wang2019neural} which tries to reverse engineer a trigger for each possible class and then use anomaly detection to predict whether the classifier is backdoored or not. 
Specifically, for one model, NC will produce an anomaly index to compare with a pre-defined threshold to determine whether a backdoor is injected into the detected model. 
Note NC will require the clean dataset for backdoor detection so we provide the whole clean test dataset to the third-party model market along with the released backdoored vanilla model $\widehat{f^C}$. 

Second, we consider STRIP~\cite{gao2019strip} by following the same assumption and setting with the authors~\cite{gao2019strip} which the correct trigger is known by the testers. 
The STRIP method will calculate the entropy value of a set of clean samples and triggered samples (samples added with correct triggers) and compare their entropy value distribution for detection. 
The clean input's entropy value calculated by STRIP is always significantly larger than the entropy of the triggered samples such that a normalized entropy distribution can determine whether a model is backdoored.  

Third, we consider DF-TND~\cite{wang2020practical} which detects if a model is backdoored without accessing datasets. 
DF-TND finds an inverted image that maximizes neuron activation, which can reveal the characteristics of the trojan signature from model weights. 
For each class, DF-TND calculates a detection score by comparing the change of logit outputs with respect to the inverted image and the random seed images. If the score of one class $K$ is larger than a pre-defined threshold, it determines the model is backdoored with the target label $K$.

\noindent\textbf{Step \textcircled{5}: Backdoor activated when victim transforms the DNN to SDN.} 
Potential victims may acquire the security-certified third-party models and transform them into multi-exit models based on the SDN structure. 
Before the victims transform this model, we assume he may use other defense schemes to modify the model to remove potential backdoors. 
We consider two backdoor removal methods, \ie, unlearning~\cite{zeng2021adversarial} and fine-tuning~\cite{li2021neural} use partially clean datasets to remove potential backdoors. 
Unlearning first finds a trigger to maximize the prediction loss for the correct label and then tries to find model parameters to make the adversarial loss given by the trigger minimized. 
Fine-tuning uses clean datasets to try to remove the effects of the backdoor injection. 
Note these backdoor removal methods introduce uncontrollable accuracy influence on models such that they are normally not used by third-party model platforms.

Besides these backdoor defense schemes, we assume the victim can acquire partial clean datasets for the IC training (\eg, partial training datasets) such that the parameters in the ICs used by the victims are clean. 
Also, when the victim transforms our released DNN models into SDN models, we assume the victim will try to add and train ICs at any reasonably hidden layers. 
After the SDN models are built, the backdoor injected will be activated to make these SDN models become vulnerable to our backdoor attacks.

\section{Experiments}
\label{sec:experiments}
In this section, we empirically validate the effectiveness and stealthiness of our attack using \textit{three} mainstream DNN architectures and \textit{four} datasets.

\subsection{Experimental setup}
\noindent\textbf{Models \& Datasets.}
We choose the following model architectures:
ResNet-56~\cite{he2016deep}, VGG-16~\cite{vgg16}, and MobileNet~\cite{howard2017mobilenets}.
The models are then transformed into SDN models for backdoor injection. 
With $p=0.8$, there are 22 ICs for ResNet-56 (a 28-layer model), 11 ICs for VGG-16 (a 14-layer model), and 12 ICs for MobileNet (a 15-layer model).
The balance parameter in \eqref{eq:final_loss} is set $\lambda=1$. 
We use datasets CIFAR-10 (C10)~\cite{CIFAR10} and Tiny-ImageNet (TI)~\cite{le2015tiny} for image classification, the dataset SVHN~\cite{svhn} for digit classification, and the dataset GTSRB~\cite{gtsrb} for traffic sign classification.
We resize the images of GTSRB to the same size as CIFAR-10, \ie, $3\times 32\times 32$. 
More details of datasets are in Table~\ref{tab:table_stat}.

\begin{table}[!htbp]
\centering
\caption{Statistics of each dataset.}
\label{tab:table_stat}
\vspace{-2ex}
\resizebox{0.99\linewidth}{!}{
\begin{tabular}{ccccc}
\Xhline{1pt}
\textbf{Dataset} & \textbf{C10} & \textbf{SVHN} & \textbf{GTSRB} & \textbf{TI}\\\Xhline{1pt}
\# of training data & 50,000 & 73,257 & 39,208 & 100,000 \\ \hline
\# of test data & 10,000  & 26,032 & 12,630 & 10,000 \\ \hline
\# of classes & 10  & 10  & 43 & 200 \\ \Xhline{1pt}

\end{tabular}
}
\end{table}

We use Pytorch 1.10 backend for the implementation. We conduct the experiments on a server equipped with two Intel Xeon 2678 V3 CPUs and 8 NVIDIA GeForce RTX 3080Ti GPUs. 
We train the DNN models using an SGD optimizer with a learning rate of 0.01 and momentum of 0.9 for 50-70 epochs in order to reach loss convergence.
Then we use the Adam optimizer~\cite{kingma2014adam} to train ICs.
We extend the training epochs of MobileNet to 100 because of its lower generalization capacity.
The parameters of transforming the DNN to the SDN are the same with~\cite{kaya2019shallow}. 

\begin{figure}[!htbp]
    \centering
    \includegraphics[width=0.8\linewidth]{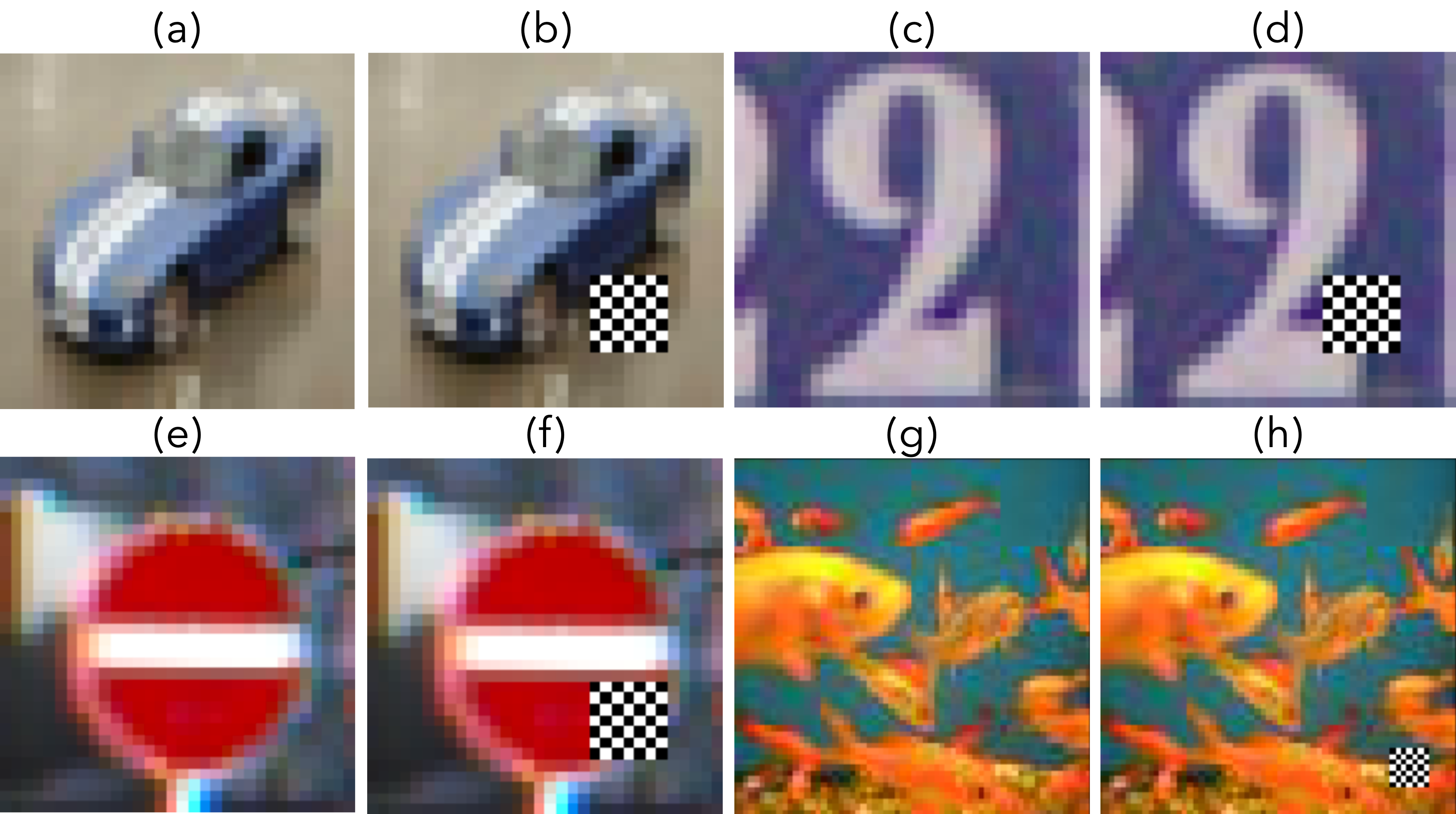}
    \caption{Examples of clean and triggered samples: (a) and (b) for CIFAR-10, (c) and (d) for SVHN, (e) and (f) for GTSRB, (g) and (h) for Tiny-ImageNet. Note that the simple trigger we use here also ensures the practicability for a physical backdoor attack. For example, attackers can stick the trigger on a traffic sign to activate the backdoor.}
    \label{fig:example}
\end{figure}

\noindent\textbf{Backdoor triggers.}
Our attack has no assumption on backdoor triggers because it does not rely on a sophisticated trigger to evade detection. 
This trigger simplicity is another strength of our attack. 
Hence, in our experiments, we adopt the checkerboard trigger of size $7\times 7$ located on the bottom-right corner of images, which is similar to the BadNets~\cite{gu2019badnets}. 

Examples are also shown in \figurename~\ref{fig:example} which represent clean samples and the samples injected by checkerboard triggers. 
Note that our attack is not related to the trigger category and similar results will be obtained for different triggers.

\noindent\textbf{Metrics.} 
We evaluate our method for effectiveness and stealthiness. 
For effectiveness, we use the Attack Success Rate (ASR)~\cite{li2020invisible,gu2019badnets,composite_backdoor} which means the proportion of successful backdoor activation among the triggered test data. 
Formally, for the attacker's target class $y_t$ and the backdoor function $\mathcal{B}$ with mask $m$ and pattern $p$, the ASR for the victim's SDN model $\widehat{f^{S'}}$ is as Eq.~\eqref{eq:def_asr}. 

\begin{equation}
    \label{eq:def_asr}
    ASR = \frac{1}{\lvert \mathcal{X}\rvert}
    \sum\limits_{(\mathbf{x}_i, y_i) \in \mathcal{X}}\mathbbm{1}(\widehat{f^{S'}}(\mathcal{B}(\mathbf{x}_i, (m,p))) = y_t),
\end{equation}
where $\lvert \mathcal{X}\rvert$ is the number of samples in the test dataset $\mathcal{X}$ and $\mathbbm{1}({A})$ equals to $1$ if event $A$ is true otherwise $0$. 

We use the accuracy on clean data (ACC) of the released backdoored vanilla model $\widehat{f^C}$ for stealthiness. 
Note that for the Tiny-ImageNet, we use the ACC top-5 while for the other three datasets we use the ACC top-1 as the metric. 

\noindent\textbf{Backdoor detection methods.} 
In this paper, we adopt three state-of-the-art detection methods including NC~\cite{wang2019neural}, DF-TND~\cite{wang2020practical}, and STRIP~\cite{gao2019strip} to detect if a backdoor exists in our released backdoored vanilla model $\widehat{f^C}$. 

For NC, we calculate the anomaly index ($I_A$) of NC for $\widehat{f^C}$. 
The implementation of NC is the same as the one proposed by the authors~\cite{wang2019neural} and the clean dataset used by NC is the clean test dataset ($\mathcal{X}_{test}$). 
The anomaly index can be calculated as: $I_A = NC(\widehat{f^C}, \mathcal{X}_{test})$.

For DF-TND, we follow the same settings in~\cite{wang2020practical} by using random noise input to find the inverted images with the pre-defined threshold as 100. 
We only list the predicted maximum score of all classes for each dataset and model.

For STRIP, we follow the same setting in~\cite{gao2019strip} which calculates the entropy distribution of tested 2,000 benign and 2,000 triggered samples for all released backdoored vanilla models $\widehat{f^C}$ of each dataset and model structure. 

\noindent\textbf{Backdoor removal methods.} 
We also evaluate two backdoor removal methods against our attack including backdoor unlearning~\cite{zeng2021adversarial} and standard fine-tuning following~\cite{li2021neural}.  

For backdoor unlearning, we follow the same setting in~\cite{zeng2021adversarial} including the ratio of clean datasets and epoch numbers (\ie, 5 epochs) to unlearn one model. 
For all architectures on CIFAR10, SVHN, and GTSRB, we further test 100 epochs (maximum epochs in~\cite{zeng2021adversarial}) and the results are similar. 
For fine-tuning, we follow the setting in~\cite{li2021neural} with 20 epochs (maximum epoch number in~\cite{li2021neural}). 

\subsection{Attack Effectiveness Evaluation}

The victims build their own SDN models by inserting ICs at their chosen hidden layers of the DNN model which is not known to the adversary. 
The IC structure and setting will follow the same as the SDN model~\cite{kaya2019shallow}. 
We assume that the victim has partial clean training data to train these ICs by themselves.  
Apparently, a victim can add as many ICs as possible to maximize the early exit ratio. 
However, ICs contain parameters such that adding ICs on layers that nearly no samples will exit will lead to unnecessary storage or training costs. 
Therefore, in this paper, we consider a scenario in which the victim will insert ICs starting from the $N_v$-th DNN hidden layer until the $n$ ($n =\lfloor pN \rfloor$) layers. 
We set $N_v > 1$ and $p = 0.8$ since very low ratio (less than 2\% for VGG-16, C10 and less than 0.1\% for VGG-16 TI) will be able to exit only based on the 1st hidden layer's and the last 20\% hidden layers' outputs.

\begin{table*}[!htbp]
\caption{The ASR for different models and datasets with different $N_v$ (2 to 7).}
\label{tab:asr}
\centering
\resizebox{0.95\linewidth}{!}{
    \begin{tabular}{c|c|c|c|c|c|c|c|c|c|c|c|c} \Xhline{1pt}
    \multirow{2}*{\textbf{$N_v$}}  & \multicolumn{4}{c|}{\textbf{ResNet-56}} & \multicolumn{4}{c|}{\textbf{VGG-16}} & \multicolumn{4}{c}{\textbf{MobileNet}} \\\cline{2-13}
    
    & \textbf{C10} & \textbf{SVHN}  & \textbf{GTSRB} & \textbf{TI} & \textbf{C10} & \textbf{SVHN} & \textbf{GTSRB} & \textbf{TI} & \textbf{C10} & \textbf{SVHN} & \textbf{GTSRB} & \textbf{TI} \\ \Xhline{1pt}
    2 & 80.1\% & 94.8\% & 85.4\% & 87.1\% & 77.9\% & 98.4\% & 90.5\% & 92.7\% & 57.8\% & 40.5\% & 52.3\% & 41.1\% \\
    3 & 83.8\% & 95.5\% & 86.6\% & 87.1\% & 91.9\% & 99.9\% & 94.7\% & 92.7\% & 61.7\% & 47.8\% & 53.3\% & 44.7\% \\
    4 & 84.3\% & 98.2\% & 86.2\% & 87.1\% & 93.7\% & 99.7\% & 94.5\% & 92.7\% & 62.5\% & 56.7\% & 54.1\% & 73.4\% \\
    5 & 84.4\% & 98.4\% & 88.7\% & 87.1\% & 95.1\% & 99.7\% & 94.5\% & 92.7\% & 68.1\% & 94.6\% & 58.1\% & 99.8\% \\
    6 & 84.6\% & 97.1\% & 89.6\% & 87.1\% & 95.9\% & 99.6\% & 94.3\% & 92.9\% & 71.6\% & 97.7\% & 62.2\% & 99.8\% \\
    7 & 84.6\% & 96.1\% & 93.4\% & 87.2\% & 95.9\% & 99.0\% & 90.5\% & 93.1\% & 79.3\% & 99.1\% & 64.2\% & 95.5\% \\\Xhline{1pt}
    \end{tabular}}
\end{table*}

The ASR for different datasets and model structures are reported in~\tablename~\ref{tab:asr} by testing triggered samples with Eq.~\eqref{eq:def_asr} of the $N_v$ ranges from 2 to 7. 
For instance, $N_v = 2$ for the ResNet-56 (a 28-layer model) model means the victim tries to insert ICs from the 2nd layer until the 22nd layer.
We notice that except MobileNet, there is certain randomness for training ICs with different datasets but our high ASR holds for experimenting with different clean datasets (partial training dataset or at least the dataset with the same distribution) as the training set. 
The reason for the lower ASR on MobileNet is that the generalization capacity of MobileNet is worse than the other two architectures.
This is also confirmed in \tablename~\ref{tab:acc_cnn} where the clean ACC of MobileNet is lower than ResNet-56 and VGG-16.
The low generalization capacity leads to the difficulty of trigger injection, thus lower ASR.
Even though, the ASR of MobileNet is still around 60\%, indicating that the attacker can successfully activate the backdoor for at least every two inferences. 

Since very few samples of the TI dataset can meet the early-exit criteria to exit from the first 6 hidden layers' outputs, we can observe different $N_v$ from 2 to 6 have the same ASR. 
Continuing to increase the $N_v$ has a tiny difference on the ASR. 
In summary, the results show that for ResNet-56 and VGG-16, our ASR is very high to illustrate the effectiveness goal of our backdoor attack.
For MobileNet, the ASR scores are lower than the other two architectures but still high enough to enable the attacker to activate the backdoor within at most 3 inputs in expectation.
We suspect the reason is rooted in the architecture difference: MobileNet learns slower than the other two architectures (\ie, requiring more epochs to converge) so it is less sensitive to triggers

\subsection{Attack Stealthiness Evaluation}
We evaluate the stealthiness of the ACC influence, the backdoor detection results, and the backdoor removal results.

\begin{table}
\centering
\caption{Stealthiness on ACC influence: ACC of clean DNN models$\rightarrow$ACC of backdoored vanilla DNN models.}
\vspace{-2mm}
\label{tab:acc_cnn}
\resizebox{0.99\linewidth}{!}{
\begin{tabular}{c|c|c|c}
\Xhline{1pt}
\textbf{Dataset} & \textbf{ResNet-56} & \textbf{VGG-16} & \textbf{MobileNet}\\ \Xhline{1pt}
C10 (top-1 ACC) & 86.9\%$\rightarrow$85.9\% & 90.8\%$\rightarrow$88.8\% & 82.7\%$\rightarrow$82.9\%\\ \hline
SVHN (top-1 ACC) & 94.8\%$\rightarrow$94.3\% & 95.3\%$\rightarrow$93.8\% & 91.3\%$\rightarrow$90.3\%\\ \hline
GTSRB (top-1 ACC) & 96.0\%$\rightarrow$95.6\% & 97.7\%$\rightarrow$97.4\% & 89.4\%$\rightarrow$87.7\%\\ \hline
TI (top-5 ACC) & 70.0\%$\rightarrow$68.3\% & 78.0\%$\rightarrow$76.9\% & 61.5\%$\rightarrow$58.2\%\\ 
\Xhline{1pt}

\end{tabular}}
\vspace{-2mm}
\end{table}

\noindent\textbf{ACC influence evaluation.} 
We release the backdoored model $\widehat{f^C}$ as a vanilla one to a third-party platform for evaluation. 
We suppose to release the test dataset as well for the third-party model platform to test this model's ACC. 
The evaluation of the ACC influence brought by our backdoor attack is given in~\tablename~\ref{tab:acc_cnn}. 
The ACC drop due to our backdoor attack is very limited (\eg, less than 2\% on C10 and TI, and less than 1\% for the other two datasets). 
Therefore, the stealthiness goal considering the ACC influence on clean samples can be achieved.

\begin{table}
\centering
\caption{Backdoored vanilla DNN models' ACC and ASR for triggered samples with correct triggers.}
\vspace{-2mm}
\label{tab:acc_backdoor1}
\resizebox{0.99\linewidth}{!}{
\begin{tabular}{c|c|c|c}
\Xhline{1pt}
\textbf{Dataset} & \textbf{ResNet-56} & \textbf{VGG-16} & \textbf{MobileNet}\\ \Xhline{1pt}
C10 (top-1 ACC/ASR) & 76.4\%/0.8\% & 67.4\%/0.1\% & 70.7\%/0.7\%\\ \hline
SVHN (top-1 ACC/ASR) & 86.2\%/1.8\% & 76.7\%/0.0\% & 74.1\%/0.0\%\\ \hline
GTSRB (top-1 ACC/ASR) & 86.0\%/0.7\% & 82.6\%/0.0\% & 65.7\%/0.1\%\\ \hline
TI (top-5 ACC/ASR) & 60.2\%/0.0\% & 72.4\%/0.5\% & 50.8\%/0.0\%\\ \Xhline{1pt}
\end{tabular}}
\vspace{-2mm}
\vspace{-2mm}
\end{table}

We also evaluate the released backdoored vanilla model's behavior for the samples with the correct triggers. 
The results are given in~\tablename~\ref{tab:acc_backdoor1}. 
We can see that adding triggers can compromise the visual contents of samples which further compromises the ACC. 
However, the ASR keeps almost zero which proves the backdoor cannot be activated for the released backdoored vanilla model.

\noindent\textbf{Backdoor detection evaluation.} 
We experiment with 3 backdoor detection methods for the released backdoored vanilla models and list the results as follows.

\tablename~\ref{tab:nc_index} shows the highest $I_A$ produced by NC for the model $\widehat{f^C}$. 
Note the pre-defined threshold of the $I_A$ for determining whether a model contains a backdoor is 2. 
The results show that the $I_A$ for all cases are less than 2 which means NC cannot detect the existence of backdoors in our published vanilla backdoored models. 
Note that NC detects backdoor per class and one of the limitations of NC is it generates huge costs for models of many classes, so we don't use NC for TI because it requires 8 GPU days for TI for one model. 
This is $50\times$ more time cost than training a model from scratch which is not realistic.
\begin{table}
\centering
\caption{Anomaly index $I_A$ of NC on backdoored vanilla DNN models (less than 2 means no backdoor detected).}
\vspace{-2mm}
\label{tab:nc_index}
\resizebox{0.7\linewidth}{!}{
\begin{tabular}{c|c|c|c}
\Xhline{1pt}
\textbf{Architecture} & \textbf{C10} & \textbf{SVHN} & \textbf{GTSRB} \\ \Xhline{1pt}
ResNet-56 & 1.032 & 0.861 & 1.826 \\ \hline
VGG-16 & 1.311 & 0.894 & 1.998 \\ \hline
MobileNet & 1.560 & 1.962 & 1.474\\ \Xhline{1pt}
\end{tabular}}
\end{table}

\begin{table}
\centering
\caption{The detection score produced by DF-TND (less than 100 means no backdoor detected).}
\vspace{-2mm}
\label{tab:dftnd_index}
\resizebox{0.8\linewidth}{!}{
\begin{tabular}{c|c|c|c|c}
\Xhline{1pt}
\textbf{Architecture} & \textbf{C10} & \textbf{SVHN} & \textbf{GTSRB} & \textbf{TI}\\ \Xhline{1pt}
ResNet-56 & 9.51 & 30.01 & 36.63 & 9.83 \\ \hline
VGG-16 & 6.14 & 6.20 & 30.27 & 25.94 \\\hline
MobileNet & 2.30 & 12.73 & 14.57 & 6.22 \\ \Xhline{1pt}
\end{tabular}}
\vspace{-4mm}
\end{table}

\begin{figure*}[t]
    \centering
    \includegraphics[width=0.95\linewidth]{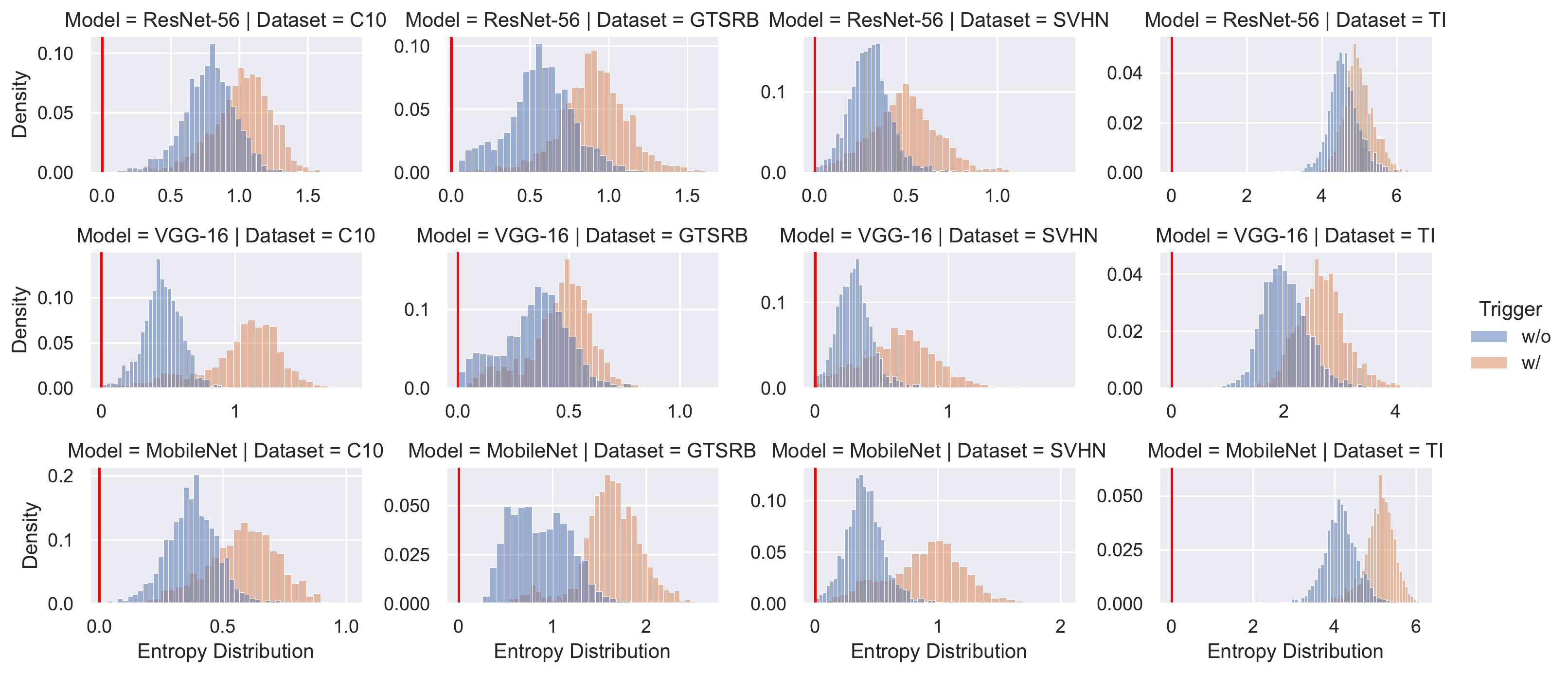}
    \caption{Entropy distribution of the backdoored DNN evaluated by STRIP. The red line is a typical detection threshold.}
    \label{fig:STRIP}
    \vspace{-5mm}
\end{figure*}

\tablename~\ref{tab:dftnd_index} shows the detection score of the DF-TND method. 
Note that for each model, DF-TND will give one detection score for each class but we only list the maximum detection score here. 
All maximum detection score is less than the detection threshold which is 100. 
This indicates that DF-TND cannot detect the existence of backdoors in our published vanilla backdoored models.

The evaluation of STRIP is in~\figurename~\ref{fig:STRIP}. 
We list the entropy distribution for the clean samples and triggered samples for comparison. 
Note a typical entropy distribution when the backdoor is detected is that all samples with triggers have low entropy values which are close to zeros. 
According to~\cite{gao2019strip}, we draw such a distribution in the left-upper subfigure in~\figurename~\ref{fig:STRIP}. 
We observe that the triggered samples can generate even larger entropy values than the clean samples. 
Thus, we can conclude that STRIP cannot detect the existence of backdoors in our published vanilla backdoored models.

\begin{table}[!htbp]
\centering
\caption{Backdoor unlearning evaluation results.}
\vspace{-2mm}
\label{tab:backdoor_unlearning}
\begin{tabular}{c|cc|cc|cc}
\Xhline{1pt}
\multirow{2}*{\textbf{Dataset}}  & \multicolumn{2}{c|}{\textbf{ResNet-56}} & \multicolumn{2}{c|}{\textbf{VGG-16}} & \multicolumn{2}{c}{\textbf{MobileNet}}\\\cline{2-7}
&\textbf{ACC}&\textbf{ASR}&\textbf{ACC}&\textbf{ASR}
&\textbf{ACC}&\textbf{ASR}\\
\Xhline{1pt}
C10&92.1\%&83.6\%&93.8\%&91.3\%&91.5\%&88.0\%\\\hline
SVHN&96.4\%&99.9\%&96.4\%&63.2\%&93.1\%&85.8\%\\\hline
GTSRB&99.1\%&84.2\%&99.7\%&78.8\%&98.8\%&46.8\%\\\hline
TI&70.0\%&66.6\%&81.9\%&60.3\%&74.7\%&98.0\%\\
\Xhline{1pt}
\end{tabular}
\vspace{-2mm}
\end{table}

\begin{table}[!htbp]
\centering
\caption{Fine-tuning evaluation results.}
\vspace{-2mm}
\label{tab:fine-tuning}
\begin{tabular}{c|cc|cc|cc}
\Xhline{1pt}
\multirow{2}*{\textbf{Dataset}}  & \multicolumn{2}{c|}{\textbf{ResNet-56}} & \multicolumn{2}{c|}{\textbf{VGG-16}} & \multicolumn{2}{c}{\textbf{MobileNet}}\\\cline{2-7}
&\textbf{ACC}&\textbf{ASR}&\textbf{ACC}&\textbf{ASR}
&\textbf{ACC}&\textbf{ASR}\\
\Xhline{1pt}
C10&85.8\%&81.3\%&89.0\%&97.9\%&82.9\%&78.7\%\\\hline
SVHN&94.4\%&97.1\%&93.9\%&99.0\%&90.3\%&76.8\%\\\hline
GTSRB&95.6\%&88.2\%&97.5\%&99.5\%&87.6\%&56.4\%\\\hline
TI&66.2\%&76.8\%&76.4\%&96.2\%&56.2\%&53.1\%\\\Xhline{1pt}
\end{tabular}
\vspace{-2mm}
\end{table}

\noindent\textbf{Backdoor removal evaluation.} 
Note that for these two backdoor removal methods, we use all the strongest settings and present the results. 
\tablename~\ref{tab:backdoor_unlearning} shows the ACC and ASR after backdoor unlearning for all datasets and models. 
Since backdoor unlearning uses a partial test set to perform unlearning and uses the rest test set to test ACC, we can observe the ACC for clean samples is improved. 
However, the ASR still remains at a high level which means it fails to remove the backdoor. 
For finetuning, we can observe that in most cases the ASR has a tiny change in~\tablename~\ref{tab:fine-tuning}.

In summary, our backdoor attack can achieve effectiveness and stealthiness for attacking dynamic multi-exit models at deployment. 
We show the possibility that a backdoor can be injected into a vanilla model for release and can evade various backdoor detection or removal methods as a security-certified third-party model for distribution. 
However, the backdoor can be effectively activated when the victims acquire such security-certified models and transform them into SDN-based multi-exit models by themselves.

\section{Discussions and Future Work}
\label{sec:discussion}

\noindent\textbf{Attack transferability.}
Our trigger injection reduces the representation gap between poisoned samples and samples of the target class in the internal layer rather than the layers of the early exits.  
Thus, even if the victim retrains ICs, the new ICs' decision boundary for the target class should be shaped by the internal representations of target class, which can encompass the representations of poisoned samples to maintain the backdoor effective. 
We note there are other multi-exits structures such as MSDNet~\cite{huang2017multi}, but as different ICs do not change the internal representations, our attack can transfer for other multi-exit architectures.
In this paper, we choose SDN for experimentation but our methodology can be extended to other dynamic multi-exit architectures. 
We will analyze how the internal decision boundary changes to understand the transferability of poisoning shallow layers for the backdoor as our future work.

\noindent\textbf{Potential countermeasure.}
Current backdoor detection methods are mainly designed for the static vanilla DNN models. 
One important assumption of these methods is that a backdoor injection will mainly influence the hidden layer of the model (\ie, the feature space~\cite{huang2022backdoor}). 
The other fact of existing backdoor attacks is that the backdoored model will misbehave with triggered samples. 
However, we only poison the shallow layers and covering up the backdoor misbehavior when releasing. 
Thus, the future countermeasure is to detect the poisoning at any layer of a model. 

\noindent\textbf{More sophisticated backdoor.}
Using different triggers for a backdoor is orthogonal to our attack. 
Thus, we pick the simple trigger design to experiment which can prove the effectiveness of our attack methodology (\ie, naive trigger can achieve high ASR and stealthiness). 
Then, using more sophisticated backdoor methods (e.g., invisible triggers~\cite{li2021invisible}) can potentially increase the ASR and bypass more future backdoor detection methods. 
We list our third future work as designing novel backdoor methods for dynamical DNNs. 

\section{Conclusion}
\label{sec:conclusion}
Numerous research works have been proposed to detect or remove backdoors in third-party DNN models. 
However, studies of backdoor attacks on the dynamic multi-exit DNNs were still not explored. 
In this work, we propose a simple yet effective backdoor attack on the dynamic multi-exit DNN models deployed in edge computing via injecting stealthy backdoors. 
Once our backdoored DNN model is transformed into a multi-exit SDN model at deployment by the victims themselves, the backdoor will be activated. 
Extensive experiments showed that our backdoor attack can not only achieve a high attack success rate but also behave normally on clean datasets and evade various backdoor defenses. 
The authors have provided public access to their code at \url{https://github.com/chichidd/BackdoorDynamicDNN}.

\section*{Acknowledgment}
We thank anonymous reviewers for their constructive comments.
We also thank Shaofeng Li for his inspiring discussion.
This work is supported by the National Key R\&D Program
of China (2022YFB3105200), NSFC under Grant No. 62106127, Singapore Ministry of Education (MOE) AcRF Tier 2 MOE-T2EP20121-0006, and Ant Group through CCF-Ant Innovative Research Program No. RF2021002.

\clearpage
\small
\bibliographystyle{IEEEtran}
\bibliography{egbib}

\begin{thebibliography}{10}
\providecommand{\url}[1]{#1}
\csname url@samestyle\endcsname
\providecommand{\newblock}{\relax}
\providecommand{\bibinfo}[2]{#2}
\providecommand{\BIBentrySTDinterwordspacing}{\spaceskip=0pt\relax}
\providecommand{\BIBentryALTinterwordstretchfactor}{4}
\providecommand{\BIBentryALTinterwordspacing}{\spaceskip=\fontdimen2\font plus
\BIBentryALTinterwordstretchfactor\fontdimen3\font minus
  \fontdimen4\font\relax}
\providecommand{\BIBforeignlanguage}[2]{{%
\expandafter\ifx\csname l@#1\endcsname\relax
\typeout{** WARNING: IEEEtran.bst: No hyphenation pattern has been}%
\typeout{** loaded for the language `#1'. Using the pattern for}%
\typeout{** the default language instead.}%
\else
\language=\csname l@#1\endcsname
\fi
#2}}
\providecommand{\BIBdecl}{\relax}
\BIBdecl

\bibitem{xu2021soda}
W.~Xu, H.~Song, L.~Hou, H.~Zheng, X.~Zhang, C.~Zhang, W.~Hu, Y.~Wang, and
  B.~Liu, ``Soda: Similar 3d object detection accelerator at network edge for
  autonomous driving,'' in \emph{IEEE INFOCOM}, 2021.

\bibitem{liu2018edge}
Q.~Liu, S.~Huang, J.~Opadere, and T.~Han, ``An edge network orchestrator for
  mobile augmented reality,'' in \emph{IEEE INFOCOM}, 2018.

\bibitem{9492000}
J.~Li, Y.~Meng, L.~Ma, S.~Du, H.~Zhu, Q.~Pei, and X.~Shen, ``A federated
  learning based privacy-preserving smart healthcare system,'' \emph{IEEE
  Transactions on Industrial Informatics}, vol.~18, no.~3, pp. 2021--2031,
  2022.

\bibitem{yu2022thwarting}
L.~Yu, S.~Zhang, L.~Zhou, Y.~Meng, S.~Du, and H.~Zhu, ``Thwarting longitudinal
  location exposure attacks in advertising ecosystem via edge computing,'' in
  \emph{IEEE ICDCS}, 2022.

\bibitem{han2018bandwidth}
S.~Han and W.~J. Dally, ``Bandwidth-efficient deep learning,'' in \emph{IEEE
  DAC}, 2018.

\bibitem{hu2019dynamic}
C.~Hu, W.~Bao, D.~Wang, and F.~Liu, ``Dynamic adaptive {DNN} surgery for
  inference acceleration on the edge,'' in \emph{IEEE INFOCOM}, 2019.

\bibitem{DBLP:conf/mobicom/LaskaridisVALL20}
S.~Laskaridis, S.~I. Venieris, M.~Almeida, I.~Leontiadis, and N.~D. Lane,
  ``{SPINN:} synergistic progressive inference of neural networks over device
  and cloud,'' in \emph{ACM MobiCom}, 2020.

\bibitem{fang2018nestdnn}
B.~Fang, X.~Zeng, and M.~Zhang, ``Nestdnn: Resource-aware multi-tenant
  on-device deep learning for continuous mobile vision,'' in \emph{Proceedings
  of the 24th Annual International Conference on Mobile Computing and
  Networking}, 2018, pp. 115--127.

\bibitem{kaya2019shallow}
Y.~Kaya, S.~Hong, and T.~Dumitras, ``Shallow-deep networks: Understanding and
  mitigating network overthinking,'' in \emph{ICML}, 2019.

\bibitem{zhou2020bert}
W.~Zhou, C.~Xu, T.~Ge, J.~McAuley, K.~Xu, and F.~Wei, ``Bert loses patience:
  Fast and robust inference with early exit,'' \emph{Advances in Neural
  Information Processing Systems}, vol.~33, pp. 18\,330--18\,341, 2020.

\bibitem{hu2020triple}
T.-K. Hu, T.~Chen, H.~Wang, and Z.~Wang, ``Triple wins: Boosting accuracy,
  robustness and efficiency together by enabling input-adaptive inference,'' in
  \emph{ICLR}, 2020.

\bibitem{li2020invisible}
S.~Li, M.~Xue, B.~Zhao, H.~Zhu, and X.~Zhang, ``Invisible backdoor attacks on
  deep neural networks via steganography and regularization,'' \emph{IEEE
  Transactions on Dependable and Secure Computing}, 2020.

\bibitem{chen2021badnl}
X.~Chen, A.~Salem, M.~Backes, S.~Ma, and Y.~Zhang, ``Badnl: Backdoor attacks
  against nlp models,'' in \emph{ICML 2021 Workshop on Adversarial Machine
  Learning}, 2021.

\bibitem{li2022backdoors}
S.~Li, T.~Dong, B.~Z.~H. Zhao, M.~Xue, S.~Du, and H.~Zhu, ``Backdoors against
  natural language processing: A review,'' \emph{IEEE Security \& Privacy},
  vol.~20, no.~05, pp. 50--59, 2022.

\bibitem{gu2019badnets}
T.~Gu, K.~Liu, B.~Dolan-Gavitt, and S.~Garg, ``Badnets: Evaluating backdooring
  attacks on deep neural networks,'' \emph{IEEE Access}, vol.~7, pp.
  47\,230--47\,244, 2019.

\bibitem{bai2021targeted}
J.~Bai, B.~Wu, Y.~Zhang, Y.~Li, Z.~Li, and S.-T. Xia, ``Targeted attack against
  deep neural networks via flipping limited weight bits,'' \emph{ICLR}, 2021.

\bibitem{wang2019neural}
B.~Wang, Y.~Yao, S.~Shan, H.~Li, B.~Viswanath, H.~Zheng, and B.~Y. Zhao,
  ``Neural cleanse: Identifying and mitigating backdoor attacks in neural
  networks,'' in \emph{IEEE S\&P}, 2019.

\bibitem{gao2019strip}
Y.~Gao, C.~Xu, D.~Wang, S.~Chen, D.~C. Ranasinghe, and S.~Nepal, ``Strip: A
  defence against trojan attacks on deep neural networks,'' in \emph{ACSAC},
  2019.

\bibitem{wang2020practical}
R.~Wang, G.~Zhang, S.~Liu, P.-Y. Chen, J.~Xiong, and M.~Wang, ``Practical
  detection of trojan neural networks: Data-limited and data-free cases,'' in
  \emph{ECCV}, 2020.

\bibitem{zeng2021adversarial}
Y.~Zeng, S.~Chen, W.~Park, Z.~M. Mao, M.~Jin, and R.~Jia, ``Adversarial
  unlearning of backdoors via implicit hypergradient,'' in \emph{ICLR}, 2022.

\bibitem{qiu2021deepsweep}
H.~Qiu, Y.~Zeng, S.~Guo, T.~Zhang, M.~Qiu, and B.~Thuraisingham, ``Deepsweep:
  An evaluation framework for mitigating {DNN} backdoor attacks using data
  augmentation,'' in \emph{Proceedings of the 2021 ACM AsiaCCS}, 2021, pp.
  363--377.

\bibitem{huang2022backdoor}
K.~Huang, Y.~Li, B.~Wu, Z.~Qin, and K.~Ren, ``Backdoor defense via decoupling
  the training process,'' in \emph{ICLR}, 2022.

\bibitem{DBLP:conf/wmcsa/LeontiadisLVL21}
I.~Leontiadis, S.~Laskaridis, S.~I. Venieris, and N.~D. Lane, ``It's always
  personal: Using early exits for efficient on-device {CNN} personalisation,''
  in \emph{Proceedings of the 22nd International Workshop on Mobile Computing
  Systems and Applications}, 2021.

\bibitem{DBLP:journals/information/PachecoBGCC21}
R.~G. Pacheco, K.~Bochie, M.~S. Gilbert, R.~S. Couto, and M.~E.~M. Campista,
  ``Towards edge computing using early-exit convolutional neural networks,''
  \emph{Inf.}, vol.~12, no.~10, p. 431, 2021.

\bibitem{DBLP:conf/eurosys/EbrahimiVGL22}
M.~Ebrahimi, A.~da~Silva~Veith, M.~Gabel, and E.~de~Lara, ``Combining {DNN}
  partitioning and early exit,'' in \emph{EdgeSys@EuroSys 2022: Proceedings of
  the 5th International Workshop on Edge Systems, Analytics and Networking,
  2022}.\hskip 1em plus 0.5em minus 0.4em\relax {ACM}, 2022, pp. 25--30.

\bibitem{DBLP:journals/corr/abs-2205-11269}
A.~Bakhtiarnia, N.~Milosevic, Q.~Zhang, D.~Bajovic, and A.~Iosifidis, ``Dynamic
  split computing for efficient deep edge intelligence,'' in \emph{ICML 2022
  Workshop on Dynamic Neural Networks}, 2022.

\bibitem{huang2017multi}
G.~Huang, D.~Chen, T.~Li, F.~Wu, L.~van~der Maaten, and K.~Q. Weinberger,
  ``Multi-scale dense networks for resource efficient image classification,''
  \emph{ICLR}, 2018.

\bibitem{qiu2020toward}
H.~Qiu, Q.~Zheng, T.~Zhang, M.~Qiu, G.~Memmi, and J.~Lu, ``Toward secure and
  efficient deep learning inference in dependable {I}o{T} systems,'' \emph{IEEE
  Internet of Things Journal}, vol.~8, no.~5, pp. 3180--3188, 2020.

\bibitem{he2019model}
Z.~He, T.~Zhang, and R.~B. Lee, ``Model inversion attacks against collaborative
  inference,'' in \emph{ACSAC}, 2019.

\bibitem{he2020attacking}
------, ``Attacking and protecting data privacy in edge--cloud collaborative
  inference systems,'' \emph{IEEE Internet of Things Journal}, vol.~8, no.~12,
  pp. 9706--9716, 2020.

\bibitem{dong2021fingerprinting}
T.~Dong, H.~Qiu, T.~Zhang, J.~Li, H.~Li, and J.~Lu, ``Fingerprinting multi-exit
  deep neural network models via inference time,'' \emph{arXiv preprint
  arXiv:2110.03175}, 2021.

\bibitem{li2022auditing}
Z.~Li, Y.~Liu, X.~He, N.~Yu, M.~Backes, and Y.~Zhang, ``Auditing membership
  leakages of multi-exit networks,'' in \emph{ACM CCS}, 2022.

\bibitem{li2021hidden}
S.~Li, H.~Liu, T.~Dong, B.~Z.~H. Zhao, M.~Xue, H.~Zhu, and J.~Lu, ``Hidden
  backdoors in human-centric language models,'' in \emph{ACM CCS}, 2021.

\bibitem{composite_backdoor}
J.~Lin, L.~Xu, Y.~Liu, and X.~Zhang, ``Composite backdoor attack for deep
  neural network by mixing existing benign features,'' in \emph{ACM CCS}, 2020.

\bibitem{DBLP:journals/corr/HanPTD15}
S.~Han, J.~Pool, J.~Tran, and W.~J. Dally, ``Learning both weights and
  connections for efficient neural networks,'' \emph{CoRR}, vol.
  abs/1506.02626, 2015.

\bibitem{DBLP:conf/cvpr/JacobKCZTHAK18}
B.~Jacob, S.~Kligys, B.~Chen, M.~Zhu, M.~Tang, A.~G. Howard, H.~Adam, and
  D.~Kalenichenko, ``Quantization and training of neural networks for efficient
  integer-arithmetic-only inference,'' in \emph{{IEEE CVPR}}, 2018.

\bibitem{compression_backdoor_icassp22}
H.~Phan, Y.~Xie, J.~Liu, Y.~Chen, and B.~Yuan, ``Invisible and efficient
  backdoor attacks for compressed deep neural networks,'' in \emph{IEEE
  ICASSP}, 2022.

\bibitem{DBLP:journals/tifs/TianSXE22}
Y.~Tian, F.~Suya, F.~Xu, and D.~Evans, ``Stealthy backdoors as compression
  artifacts,'' \emph{{IEEE} Trans. Inf. Forensics Secur.}, vol.~17, pp.
  1372--1387, 2022.

\bibitem{DBLP:journals/corr/abs-2108-09187}
H.~Ma, H.~Qiu, Y.~Gao, Z.~Zhang, A.~Abuadbba, M.~Xue, A.~Fu, Z.~Jiliang, S.~F.
  Al{-}Sarawi, and D.~Abbott, ``Quantization backdoors to deep learning
  commercial frameworks,'' \emph{CoRR}, vol. abs/2108.09187, 2021.

\bibitem{li2021neural}
Y.~Li, X.~Lyu, N.~Koren, L.~Lyu, B.~Li, and X.~Ma, ``Neural attention
  distillation: Erasing backdoor triggers from deep neural networks,''
  \emph{ICLR}, 2021.

\bibitem{he2016deep}
K.~He, X.~Zhang, S.~Ren, and J.~Sun, ``Deep residual learning for image
  recognition,'' in \emph{IEEE CVPR}, 2016.

\bibitem{vgg16}
K.~Simonyan and A.~Zisserman, ``Very deep convolutional networks for
  large-scale image recognition,'' \emph{arXiv preprint arXiv:1409.1556}, 2014.

\bibitem{howard2017mobilenets}
A.~G. Howard, M.~Zhu, B.~Chen, D.~Kalenichenko, W.~Wang, T.~Weyand,
  M.~Andreetto, and H.~Adam, ``Mobilenets: Efficient convolutional neural
  networks for mobile vision applications,'' \emph{arXiv preprint
  arXiv:1704.04861}, 2017.

\bibitem{CIFAR10}
A.~Krizhevsky and G.~Hinton, ``Learning multiple layers of features from tiny
  images,'' Citeseer, Tech. Rep., 2009.

\bibitem{le2015tiny}
Y.~Le and X.~Yang, ``Tiny imagenet visual recognition challenge,'' \emph{CS
  231N}, vol.~7, no.~7, p.~3, 2015.

\bibitem{svhn}
Y.~Netzer, T.~Wang, A.~Coates, A.~Bissacco, B.~Wu, and A.~Y. Ng, ``Reading
  digits in natural images with unsupervised feature learning,'' in \emph{NIPS
  Workshop on Deep Learning and Unsupervised Feature Learning}, 2011.

\bibitem{gtsrb}
S.~Houben, J.~Stallkamp, J.~Salmen, M.~Schlipsing, and C.~Igel, ``Detection of
  traffic signs in real-world images: The german traffic sign detection
  benchmark,'' in \emph{IJCNN}, 2013.

\bibitem{kingma2014adam}
D.~P. Kingma and J.~Ba, ``Adam: A method for stochastic optimization,''
  \emph{arXiv preprint arXiv:1412.6980}, 2014.

\bibitem{li2021invisible}
Y.~Li, Y.~Li, B.~Wu, L.~Li, R.~He, and S.~Lyu, ``Invisible backdoor attack with
  sample-specific triggers,'' in \emph{Proceedings of the IEEE/CVF CVPR}, 2021,
  pp. 16\,463--16\,472.

\end{thebibliography}
\end{document}